\documentclass{IEEEcsmag}

\usepackage[colorlinks,urlcolor=blue,linkcolor=blue,citecolor=blue]{hyperref}

\usepackage[disable]{todonotes}

\newcommand*{\nolink}[1]{%
  \begin{NoHyper}#1\end{NoHyper}%
}

\newcommand\notype[1]{\unskip}

\usepackage{upmath}

\usepackage{listings}
\usepackage{xcolor}

\definecolor{codegreen}{rgb}{0,0.6,0}
\definecolor{codegray}{rgb}{0.5,0.5,0.5}
\definecolor{codepurple}{rgb}{0.58,0,0.82}
\definecolor{backcolour}{rgb}{0.95,0.95,0.92}

\lstdefinestyle{mystyle}{
    backgroundcolor=\color{backcolour},   
    commentstyle=\color{codegreen},
    keywordstyle=\color{magenta},
    numberstyle=\tiny\color{codegray},
    stringstyle=\color{codepurple},
    basicstyle=\ttfamily\footnotesize,
    breakatwhitespace=false,         
    breaklines=true,                 
    captionpos=b,                    
    keepspaces=true,                 
    numbers=left,                    
    numbersep=5pt,                  
    showspaces=false,                
    showstringspaces=false,
    showtabs=false,                  
    tabsize=2
}

\jvol{XX}
\jnum{XX}
\paper{8}
\jmonth{Nov/Dec}
\jname{IT Professional}
\pubyear{2022}

\begin{document}


\title{Collaboration of Digital Twins through Linked Open Data: Architecture with FIWARE as Enabling Technology}

\author{Javier Conde}
\affil{Departamento de Ingeniería de Sistemas Telemáticos, Escuela Técnica Superior de Ingenieros de Telecomunicación, Universidad Politécnica de Madrid}
\author{Andres Munoz-Arcentales}
\affil{Departamento de Ingeniería de Sistemas Telemáticos, Escuela Técnica Superior de Ingenieros de Telecomunicación, Universidad Politécnica de Madrid}
\author{Álvaro Alonso}
\affil{Departamento de Ingeniería de Sistemas Telemáticos, Escuela Técnica Superior de Ingenieros de Telecomunicación, Universidad Politécnica de Madrid}
\author{Gabriel Huecas}
\affil{Departamento de Ingeniería de Sistemas Telemáticos, Escuela Técnica Superior de Ingenieros de Telecomunicación, Universidad Politécnica de Madrid}
\author{Joaqu\'in Salvachúa}
\affil{Departamento de Ingeniería de Sistemas Telemáticos, Escuela Técnica Superior de Ingenieros de Telecomunicación, Universidad Politécnica de Madrid}


\begin{abstract}
The collaboration of the real world and the virtual world, known as Digital Twin, has become a trend with numerous successful use cases. However, there are challenges mentioned in the literature that must be addressed. One of the most important issues is the difficulty of collaboration of Digital Twins due to the lack of standardization in their implementation. This article continues a previous work that proposed a generic architecture based on the FIWARE components to build Digital Twins in any field. Our work proposes the use of Linked Open Data as a mechanism to facilitate the communication of Digital Twins. We validate our proposal with a use case of an urban Digital Twin that collaborates with a parking Digital Twin. We conclude that Linked Open Data in combination with the FIWARE ecosystem is a real reference option to deploy Digital Twins and to enable the collaboration between Digital Twins.       
\end{abstract}

\maketitle

\chapterinitial{The evolution of technologies} such as Cloud Computing, Internet of Things (IoT), Big Data, or Artificial Intelligence (AI) have enabled the development and implementation of Digital Twins (DTs). DTs emerged in 2003 when Grieves and Vickers presented them as virtual representations of physical products \cite{grieves2014digital}. Monitoring the physical entities allow to build the history of the real world in databases, and thanks to Big Data, all this information can be processed and decisions can be made by applying AI techniques~\cite{An_architecture_of_an_Intelligent_Digital}.

DTs can be applied at all stages of the life cycle of a product, from its initial design to its retirement \cite{review_of_digital_twin_about}. In addition, they can be used in different fields, such as health (\cite{on_the_integration_of_agents}); manufacturing (\cite{the_use_of_digital_twin_for_predictive}); farming (\cite{Smart_Livestock_Farms_Using_Digital}); etc.

However, although DTs have been validated within multiple use cases, there are still challenges related to them. The main barrier is the lack of standardization which difficulties the collaboration of DTs. It means that each specific case proposes a particular solution with its own architecture. The most widespread proposal is a layer-based architecture oriented to the flow of data between the virtual world and the physical one \cite{A_Six_Layer_Digital_Twin}. These are data-oriented architectures focused on data collection, data modeling, data storage, information processing, and updating of the physical world with the results obtained from the virtual counterpart. Data modeling determines how information will be represented in the DT. Modeling is essential not only for the operation of the DT itself but also for the collaboration of DTs and the connection of external systems to the DT.

The research conducted by Conde et al. \cite{modeling_digital_twin} presents an architecture based on the FIWARE technology that is agnostic to the DT use case. FIWARE (\url{https://www.fiware.org/}) is an European initiative that provides a set of software components, called Generic Enablers (GEs), for the development of intelligent solutions. All FIWARE GEs are compliant with the ETSI Next Generation Service Interface (NGSI) standard in its two versions: NGSIv2 and NGSI-LD for Linked Data. The NGSI standard API models the information context as entities and defines the way to access these entities through the HTTP protocol \cite{enabling_Context_Aware_Data}. NGSI is domain-independent, that is, it only defines the API for accessing entities. As a result, the FIWARE ecosystem also provides the Smart Data Models initiative (\url{https://smartdatamodels.org/}) that aims to standardize data models. Smart Data Models are reference models, validated by the industry, and compatible with NGSI, that guarantee the interoperability among systems by homogenizing the way in which data are modeled.

Three years after Grieves and Vickers introduced DTs, Tim Berners-Lee coined the term Linked Data (LD) defined as a method of publishing data on the Web, by making data accessible through HTTP URIs, and in which data are interconnected with other data through URIs \cite{linked_data_the_story}. The goal of Tim Berners-Lee was to create a Semantic Web based on LD, in contrast to the original Web based on documents (Web pages). Afterwards, Tim Berners-Lee presented Linked Open Data (LOD), defined as LD that are published in a free and accessible way \cite{linked_data}.

Some researches propose Open Data (OD) to feed the DT. The work of Morgan and Ali \cite{Industrial_IoT_and_Digital_Twins} presents a DT for industrial machines which is managed by OD provided by Microsoft. Ruohomäki et al. \cite{Smart_City_Platform_Enabling_Digital_Twin} propose the implementation of an urban DT in Helsinki that combines real-time information with information from Open Data Portals (ODPs).  

The aim of our research is to solve the problem of communication of DTs. We have extended the reference architecture presented by Conde et al. by proposing  the use of LOD in combination with the FIWARE ecosystem and FIWARE Smart Data models. We validate the proposal with a use case of a DT that publishes its data in an ODP and another DT which consumes these OD and processes them as part of its operation.

The remaining of this article is structured as follows. Next section presents the integration of LOD in DTs. It is followed by the extension with LOD of the architecture based on the FIWARE GEs and the FIWARE Smart Data Models. The proposal includes how to publish LOD from a DT and how to consume LOD in a DT. Thereafter, the communication of DTs is validated in a use case of an urban DT that consumes LOD generated by a parking DT. Lastly, conclusions of the research are presented and an outline of future work is given.

\section{DT MANAGED BY LOD}

As mentioned in the introduction, some researchers propose LOD as source of information for DTs \cite{Industrial_IoT_and_Digital_Twins, Smart_City_Platform_Enabling_Digital_Twin}. However, DTs also generate a large amount of data that can be released as LOD and reused by other systems or DTs. 

The most common way to publish LOD is through ODPs \cite{realizing_the_innovation_potentials}. \textbf{CKAN} (\url{https://ckan.org/}) is an open-source software with the necessary tools to deploy an ODP. It is widely spread and used by many institutions. The CKAN API provides the methods to create organizations (i.e., the owners of the data), datasets (i.e., units of data composed of resources and metadata), and resources (i.e., the data itself). CKAN also manages the metadata needed to search for datasets. The metadata are stored in CKAN, and the data can be stored in CKAN as well, or in the original source. If CKAN does not store the data, the metadata will indicate where to find the original source. The basic functionality of CKAN can be expanded with extensions, such as DCAT CKAN (\url{https://extensions.ckan.org/extension/dcat/}). This extension adapt metadata in CKAN to be compliant with DCAT (the catalog vocabulary developed by W3C [\url{https://www.w3.org/TR/vocab-dcat-3/}]).  

In the LOD process, different stakeholders are involved. The most relevant are publishers, consumers, and ODP maintainers. The LOD life-cycle begins when publishers gather data sources that they want to publish (e.g., temperature measurement from an IoT device). They must generate the metadata, clean the data, and represent them in an accessible format for people and machines (e.g., JSON-LD). Maintainers must provide the publisher with mechanisms to publish the data in the ODP. Finally, consumers will discover datasets and exploit them \cite{A_systematic_review_of_open_government_data}. In the DT case, the DT will act as a publisher and/or consumer. From the publisher's point of view, DTs may publish their data as LOD. From the consumer's point of view, DTs can use LOD as additional data sources. Formats such as NGSI-LD and initiatives as the Smart Data Models allow the standardization of data, which is required for the communication of different systems. Standards such as DCAT for the metadata guarantees the integration of systems by helping DTs publish, discover, and consume new data sources.

\section{EXTENDING THE FIWARE DT ARCHITECTURE WITH LOD}

Our article extends the architecture proposed by Conde et al. \cite{modeling_digital_twin} in which the FIWARE ecosystem is defined to deploy DTs. The original architecture proposes the \textbf{Context Broker (CB)} (\url{https://github.com/FIWARE/context.Orion-LD}) to manage context information using the NGSI-LD standard. The CB supports two methods of connection: (1) a synchronous connection through HTTP requests; and (2) an asynchronous connection based on the publish-subscribe pattern. Information in a DT proceeds from multiple sources including IoT devices. The \textbf{IoT Agent GE} (\url{https://github.com/FIWARE/catalogue/blob/master/iot-agents/README.md}) acts as a link between the IoT devices and the CB adapting sensor measurements into NGSI-LD entities and vice versa. Other data sources can be integrated into the DT using the \textbf{Draco GE} (\url{https://fiware-draco.readthedocs.io/}). Draco is a flow-based programming system based on Apache NiFi with a set of processors compatible with the NGSI standard. Draco is used to obtain data from external sources, make them compliant with a Smart Data Model, and save them in the CB. The CB only saves the current state of the entities. Thus, Draco is needed to build the historical data of entities. With these data, Machine Learning models can be trained using the \textbf{Cosmos GE} (\url{https://fiware-cosmos.readthedocs.io/en/latest/}). The security of the DT is managed by the Keyrock, Wilma, AuthzForce GEs, and the OAuth2.0 protocol. \textbf{Keyrock} (\url{https://fiware-idm.readthedocs.io/}) is an identity manager acting as an Oauth2.0 server in charge of authentication and authorization; \textbf{Wilma} (\url{https://fiware-pep-proxy.readthedocs.io/}) is a proxy that acts as an access control point used for securing microservices, (e.g., the CB). When Wilma receives a petition, it validates the permission through Keyrock which allows simple authorization policies based on HTTP verb and path. \textbf{AuthzForce} (\url{https://authzforce-ce-fiware.readthedocs.io/}) allows more complex authorization policies specified using XACML.  

As we mentioned, in DTs a lot of data are generated and they can be included in ODPs. Consequently, the extended architecture proposes LOD as a mechanism for the collaboration of DTs. We will explain both approaches: (1) publication of LOD with the information generated in the DT; and (2) consumption of LOD to be used in the DT. For both cases, Draco will be used as the enabling technology to integrate LOD into DTs. \nolink{\textbf{Table~\ref{table:original_arch}}} summarizes the extended architecture for DTs based on the FIWARE GEs.

\subsection{Publication of LOD generated by the DT}

The DT can act as a source of LOD, however, it is necessary to extend the reference architecture based on FIWARE GEs. FIWARE Draco includes two processors to generate the metadata in CKAN format and to publish the NGSI entities and metadata in CKAN. The \texttt{UpdateCKANMetadata} processor is responsible for generating the metadata according to the information extracted from the entities. The generated metadata are compatible with the DCAT catalog vocabulary. The \texttt{NGSIToCKAN} processor generates organizations, datasets, and resources from the NGSI-LD entities. For this purpose, Draco subscribes to changes in entities and receives notifications when they are created or updated. Through these notifications, \texttt{NGSIToCKAN} publishes the data and metadata in a CKAN instance. The processor has the option of registering only the metadata. In this case, the data will be accesible directly through the CB. For example, when a \texttt{Parking} entity is created, Draco receives a notification. The \texttt{UpdateCKANMetadata} processor extracts the \texttt{id} property and configures it as the dataset name. Then, \texttt{NGSIToCKAN} creates the new dataset in the ODP using the CKAN API. The Smart Data Models offer the possibility to model the metadata of entities through the DCAT data models (\url{https://github.com/smart-data-models/dataModel.STAT-DCAT-AP}). In this case, Draco subscribes to metadata entities, extracts the information, and stores them in CKAN.

\subsection{Consumption of LOD in the DT}

In addition to publishing the data as LOD, DTs can also receive data from ODPs. In this case, the DT will act as a consumer. Metadata allow to find the desired dataset in the ODP. If the downloaded data are not compatible with the NGSI-LD standard, Draco can be used for transforming them. In the same way, if the entities do not fit any Smart Data Model, Draco can adapt them. Once the data are homogenized, it is also possible to republish them in CKAN, as a new dataset that now complies with the standard.

\begin{table}
\caption{Extended DT architecture based on FIWARE GEs.}
\label{table:original_arch}
\small
\begin{tabular*}{17.5pc}{@{}|p{50pt}<{\raggedright}|p{135pt}<{\raggedright}|@{}}
\hline
\textbf{FIWARE GE}& \textbf{Role in DT architecture} \\ 
\hline
Context Broker& $^{\mathrm{a}}$Management of NGSI-LD context information. \\ 
\hline
IoT Agent& $^{\mathrm{a}}$Integration of IoT devices into the FIWARE DT architecture. \\
\hline
Draco& $^{\mathrm{a}}$Homogenizing data sources to be compliant with NGSI-LD and the Smart Data Models. \\
& $^{\mathrm{a}}$Connecting external sources to the DT architecture. \\ 
& $^{\mathrm{a}}$Building the historical data. \\
& $^{\mathrm{b}}$Generating the metadata to be published in CKAN ODPs. \\
& $^{\mathrm{b}}$Publishing the data and metadata in CKAN ODPs. \\
& $^{\mathrm{b}}$Consuming datasets from ODPs. \\
\hline
Cosmos& $^{\mathrm{a}}$Big Data processing. \\
\hline
Keyrock& $^{\mathrm{a}}$Autentication and authorization. \\
\hline
Wilma& $^{\mathrm{a}}$Policy Enforcement Point. \\
\hline
AuthzForce& $^{\mathrm{a}}$Policy Decision Point. \\
\hline
\multicolumn{2}{@{}p{17.5pc}@{}}{$^{\mathrm{a}}$: Original DT architecture based on FIWARE GEs \cite{modeling_digital_twin}.}\\
\multicolumn{2}{@{}p{17.5pc}@{}}{$^{\mathrm{b}}$: Extended architecture to integrate LOD in DTs.}
\end{tabular*}
\end{table}

\section{USE CASE OF COLLABORATION OF DTS THROUGH LOD}

\begin{figure*}
\centerline{\includegraphics[width=37pc]{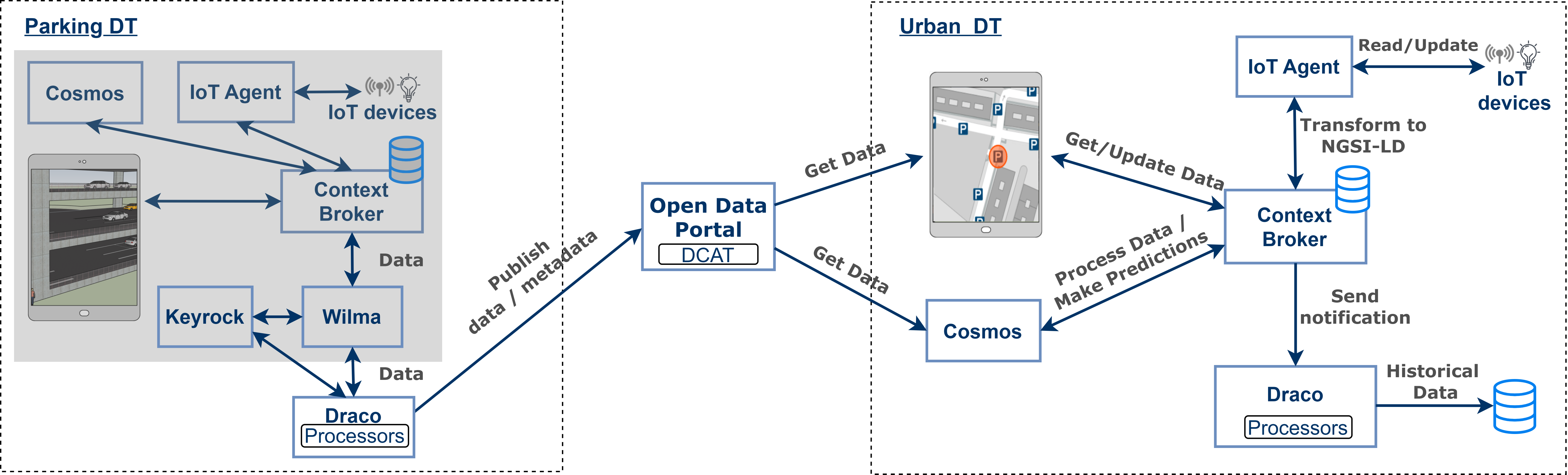}}
\caption{Architecture of the parking DT and the urban DT collaboration through LOD. The original parking DT (in gray) is the use case presented in \cite{modeling_digital_twin}.}
\label{fig:architecture}
\end{figure*}

This section presents a use case for the collaboration of two DTs through LOD. An NGSI-LD version of the parking DT presented in \cite{modeling_digital_twin} is extended with an urban DT that improves traffic in a city. This urban DT will use the LOD published by the parking DT as a data source to manage the off-street parking information. It will also receive data from IoT sensors located in the on-street parkings. The DT is completed with a mobile application that helps citizens find parking in real time. In \nolink{\textbf{Figure~\ref{fig:architecture}}} the extended architecture with the communication of both DTs is presented.

\subsection{Publication of LOD from the parking DT}

In the original parking DT the information is modeled through entities of type \texttt{OffStreetParking} (representation of an off-street parking), \texttt{ParkingSpot} (representation of a parking spot), and \texttt{Vehicle} (representation of the vehicle occupying a parking spot). Only entities of type \texttt{OffStreetParking} will be published as LOD. An example of a simplified version of an \texttt{OffStreetParking} entity is shown below:

\begin{lstlisting}
{
    "id": "urn:ngsi-ld:OffStreetParking:1",
    "type": "OffStreetParking",
    "location": { 
        "coordinates": [40.3312618, -3.7574926],
        "type": "Point"
    },
    "availableSpotNumber": 132,
    "@context": [
        "https://raw.githubusercontent.com/smart-data-models/dataModel.Parking/master/context.jsonld"
    ]
}
\end{lstlisting}

A Draco instance subscribes to changes in the \texttt{availableSpotNumber} attribute to be notified every time the number of free spots is updated. This notification is forwarded to the \texttt{NGSIToCKAN} processor which publishes the new information in a dataset of an ODP. The first time Draco receives a notification of a new off-street parking, the \texttt{UpdateCKANMetadata} and \texttt{NGSIToCKAN} processors create the corresponding organization, dataset, and resources in the ODP. The metadata generated, DCAT compliant, ensure reusability, contextuality, findability, and interoperability of the dataset. The CB is secured with Wilma and Keyrock, limiting Draco access to \texttt{OffStreetParking} entities.  

\begin{lstlisting}
<?xml version="1.0" encoding="utf-8"?>
<rdf:RDF 
 xmlns:rdf="http://www.w3.org/1999/02/22-rdf-syntax-ns#" 
 xmlns:dcat="http://www.w3.org/ns/dcat#" 
 xmlns:dct="http://purl.org/dc/terms/">
 <dcat:Dataset>
  <dct:title>
   Parking 1
  </dct:title>
  <dcat:distribution>
   <dcat:Distribution>
    <dct:title>
     Occupancy level of Parking 1
    </dct:title>
    ...
   </dcat:Distribution>
  </dcat:distribution>
 ...
 </dcat:Dataset>
</rdf:RDF>
\end{lstlisting}

\subsection{Consumption of LOD from the urban DT}
On the one hand, the new urban DT will access information from IoT sensors installed in parking lots on the street. On the other hand, it will obtain information from the off-street parkings published by the parking DT in the ODP. 

The FIWARE IoT Agents will transform the data coming from the IoT sensors into NGSI-LD entities and publishes them in the CB as \texttt{ParkingSpot} entities:

\begin{lstlisting}
{
    "id": "urn:ngsi-ld:ParkingSpot:123",
    "type": "ParkingSpot",
    "location": {
        "coordinates": [40.405382, -3.6734942],
        "type": "Point"
    },
    "status": "closed",
    "@context": [
        "https://raw.githubusercontent.com/smart-data-models/dataModel.Parking/master/context.jsonld"
    ]
}
\end{lstlisting}

In parallel, the client application will connect to the CB and the ODP to obtain a real-time representation of the city. The client application will create in the CB an entity of type \texttt{RequestParking} with the user's position to be processed by Cosmos:

\begin{lstlisting}
{
    "id": "urn:ngsi-ld:RequestParking:12345",
    "type": "RequestParking",
    "location": {
        "coordinates": [40.331262, -3.757495],
        "type": "Point"
    },
    "@context": [
        "https://uri.etsi.org/ngsi-ld/v1/ngsi-ld-core-context.jsonld"
    ]
}
\end{lstlisting}

Cosmos will receive a notification and, using the data from the on-street parkings and of-street parkings, will obtain the closest position where the user can park. Once the result is obtained, it will update a \texttt{ResponseParking} entity with the response. Finally, the client application will notify the user with the result.

Moreover, Draco can be used to build a historical database of the city. Machine Learning models can be trained using these historical data, and Cosmos can be used to make predictions. Based on the results, the DT can make decisions and modify the real world, for example, by opening the \texttt{ParkingSpot} with id 123.

\subsection{Results and limitations of the use case}

The solution presented can be adapted to Big Data scenarios. Orion can scale up horizontally through the sharding of MongoDB. Draco and Cosmos are based on Apache Nifi, Apache Spark, and Apache Flink (tools designed for Big Data processing tasks).

On the other hand, feeding the DT with LOD has limitations. Data publishers can change the data, change their format, or stop offering them. Consequently, the election of datasets results an important task. ODPs offer mechanisms to filter datasets based on their metadata (e.g., SPARQL search engines). However, in many cases these metadata do not adapt the DCAT standard or do not have enough quality. 

The architecture presented can be extended with other GEs, as Idra (\url{https://idra.readthedocs.io/}), a federator of OD sources such as CKAN ODPs or web pages; or the Scorpio GE (\url{https://scorpio.readthedocs.io/}), a CB with federation capabilities and scalability through Kafka.

\section{CONCLUSIONS AND FUTURE WORK}
DTs have experienced great growth in recent years. However, there are still challenges related to them like collaboration among DTs and their integration with external systems. This article extends the architecture proposed in a previous work based on the FIWARE \cite{modeling_digital_twin} by adding LOD as a mechanism to enable the communication of DTs. Two approaches have been analyzed: (1) DTs that publish LOD in ODPs; and (2) DTs that consume LOD from ODPs. The theoretical proposal is completed with a use case of a parking DT that collaborates with an urban DT. From the results obtained, we can conclude that LOD in combination with the FIWARE ecosystem is a legitimate reference option to deploy DTs and to make the collaboration between DTs effective.
For future research, the proposal should be validated in other scenarios. Issues related to the quality of the metadata and data should be addressed, as well as the limitations of publishing the data of a DT as open. Finally, it would be interesting to study scenarios and methods for private data that require pre-processing before being published in ODPs.

\bibliographystyle{IEEEtran}
\bibliography{IEEEabrv, ref}

\begin{IEEEbiography}{Javier Conde}{\,} is currently a researcher in the Department of Telematics Engineering. He is interested in the fields of Digital Twins, Linked Open Data, and Artifficial Intellgence.
\end{IEEEbiography}

\begin{IEEEbiography}{Andres Munoz-Arcentales} is a Ph.D. in Telematics Engineering at UPM and his research interests include Machine Learning, Data Fusion, and Big Data.
\end{IEEEbiography}

\begin{IEEEbiography}{\'Alvaro Alonso}{\,} is currently an Associate Professor with the UPM. His main research interests are Security Management, Multi-videoconferencing Systems, and Open Data.
\end{IEEEbiography}

\begin {IEEEbiography}{Gabriel Huecas}{\,} is currently an Associate Professor with the UPM. his research interests include Digitization, Big Data, and Cloud Computing.
\end{IEEEbiography}

\begin{IEEEbiography}{Joaqu\'in Salvach\'ua}{\,} is currently an Associated Professor with UPM. His research interests lie in the fields of Big Data, Data Privacy and usage control, Cloud architectures, and Blockchain.
\end{IEEEbiography}

\end{document}